\documentclass[
 reprint,
 superscriptaddress,
 amsmath,amssymb,
 floatfix
]{revtex4-1}

\usepackage[english]{babel}
\usepackage{graphicx}
\usepackage{upgreek}
\usepackage{bm}
\usepackage{textcomp}
\usepackage[cmyk,dvipsnames]{xcolor}
\usepackage[colorlinks = true,
						citecolor = blue,
						linkcolor = red,
						urlcolor = blue]{hyperref}
						
\definecolor{pacificb}{HTML}{1CA9C9}

\begin{document}

\title{Mechanisms of skyrmion collapse revealed by sub-nm maps of the transition rate}
\author{Florian Muckel}
\thanks{These authors contributed equally to this work:\\
muckel@physik.rwth-aachen.de\\
malottki@physik.uni-kiel.de}
\affiliation{II. Institute of Physics B and JARA-FIT, RWTH Aachen University, D-52074 Aachen, Germany}
\author{Stephan von Malottki}
\thanks{These authors contributed equally to this work:\\
muckel@physik.rwth-aachen.de\\
malottki@physik.uni-kiel.de}
\affiliation{Institute of Theoretical Physics and Astrophysics, University of Kiel, Leibnizstrasse 15, 24098 Kiel, Germany}
\author{Christian Holl}
\author{Benjamin Pestka}
\author{Marco Pratzer}
\affiliation{II. Institute of Physics B and JARA-FIT, RWTH Aachen University, D-52074 Aachen, Germany}

\author{Pavel F. Bessarab}
\affiliation{Science Institute of the University of Iceland, 107 Reykjav\'ik, Iceland}
\affiliation{ITMO University, 197101 St. Petersburg, Russia}
\affiliation{Peter Gr\"unberg Institute and Institute for Advanced Simulation, Forschungszentrum J\"ulich, 52425 J\"ulich, Germany}

\author{Stefan Heinze}
\affiliation{Institute of Theoretical Physics and Astrophysics, University of Kiel, Leibnizstrasse 15, 24098 Kiel, Germany}
\author{Markus Morgenstern}
\affiliation{II. Institute of Physics B and JARA-FIT, RWTH Aachen University, D-52074 Aachen, Germany}



\maketitle

\textbf{Magnetic skyrmions are key candidates for novel memory, logic, and neuromorphic computing.
An essential property is their
topological protection caused by the whirling spin texture as described by a robust integer winding number. However, the realization on an atomic lattice leaves a loophole for switching the winding number via concerted rotation of individual spins. Hence, understanding the unwinding  
microscopically is key to 
enhance skyrmion stability. Here, we use spin polarized scanning tunneling microscopy to probe 
skyrmion annihilation by individual hot electrons
and obtain maps of the transition rate on the nanometer scale.
By applying an in-plane magnetic field, we tune the collapse rate by up to four orders of magnitude.
In comparison with first-principles based atomistic spin 
simulations, the experiments demonstrate 
a radial symmetric collapse at zero in-plane magnetic field 
and a transition to
the recently predicted 
chimera collapse at finite in-plane field.
Our work opens the route to design criteria for skyrmion switches and improved skyrmion stability.}

The topological protection of skyrmions \cite{Bogdanov1989,Muhlbauer2009,Yu2010} implies large stability during manipulation \cite{Fert2017,Fert2013}. The additional possibility to tune magnetic skyrmions up to room temperature by stacking ultrathin transition-metal films \cite{Heinze2011,Jiang2015,Woo2016} as well as to control skyrmion creation and subsequent transport at relatively low current density \cite{Jiang2015,Legrand2017} established them as key candidates for race-track memories or logic devices \cite{Fert2017,Fert2013, Zhang2015} 
and prospective for synaptic network electronics \cite{Song2020}.
However, the stability of the skyrmions, not rigorously protected by topology  \cite{Abanov1998,Cai2012}, is a key challenge 
\cite{Milde2013,Kagawa2017,Je2020}.  While a small skyrmion size is mandatory for favorable storage density and energy efficiency \cite{Fert2017, Fert2013}, the energy barrier preventing skyrmion collapse tends to strongly decrease with reduced skyrmion size~\cite{Cai2012,Varentsova2018}.
Experimentally, intriguing results have been found regarding the Arrhenius prefactor of the skyrmion collapse rate that varies by 30 orders of magnitude with out-of-plane magnetic field \cite{Wild2017} and showcases a strong dependence on local disorder \cite{Zzvorka2019,Legrand2017,Woo2018}. The central handle to probe the skyrmion dynamics is real space  mapping  \cite{Jiang2015,Jiang2016,Boulle2016,Woo2016,Soumyanarayanan2017,MoreauLuchaire2016,Chen2015,Yu2010,Seki2012,Park2014,Milde2013,Dovzhenko2018,Grenz2017,Heinze2011,Romming2013,Romming2015, Hanneken2015,Herv2018,Meyer2019}. It has been employed to reveal controlled writing and deleting \cite{Romming2013,Hsu2016}, creation at defects via spin currents \cite{Jiang2015,Buettner2017}, current induced longitudinal and transversal motion \cite{Jiang2015,Woo2016,Jiang2016,Litzius2016} as well as for pinpointing excitation modes \cite{Buettner2015,Woo2017}.  
The most advanced method for mapping thin film skyrmions is spin-polarized scanning tunneling microscopy (SP-STM) \cite{Heinze2011,Romming2013,Romming2015, Hanneken2015,Herv2018,Meyer2019}. It provided guiding insights by employing the paradigmatic Pd/Fe bilayer on Ir(111)
\cite{Romming2013,Romming2015,Hagemeister2015,Hanneken2015,Kubetzka2017}. The system exhibits ultra-small, isolated 
N\'eel skyrmions \cite{Romming2013,Romming2015,Kubetzka2017} that are often pinned at atomic defects within the Pd layer \cite{Romming2013,Hanneken2016}. Its current-induced collapse has been probed \cite{Romming2013,Hagemeister2015} and subsequently compared to Monte Carlo simulations (MCS) without pinpointing to a mechanism \cite{Hagemeister2015}. 

Independently of such experiments, atomistic spin simulations \cite{Eriksson2017} based on parameters from density functional theory (DFT) found a strong entropy contribution to the collapse prefactor \cite{vonMalottki2019,Bessarab2018} and key contributions to the energy barrier by exchange frustration \cite{vonMalottki2017}, higher-order exchange interactions \cite{Paul2019} or the presence of defects \cite{Uzdin2018}.
In addition, a novel collapse mechanism besides the conventionally assumed radial symmetric collapse has been predicted~\cite{Meyer2019,Desplat2019}, 
coined the chimera mode due to the intermediate state with 
an unconventional dipolar topological charge. 
\\    

\begin{figure*}[ht]
\includegraphics[width=176mm]{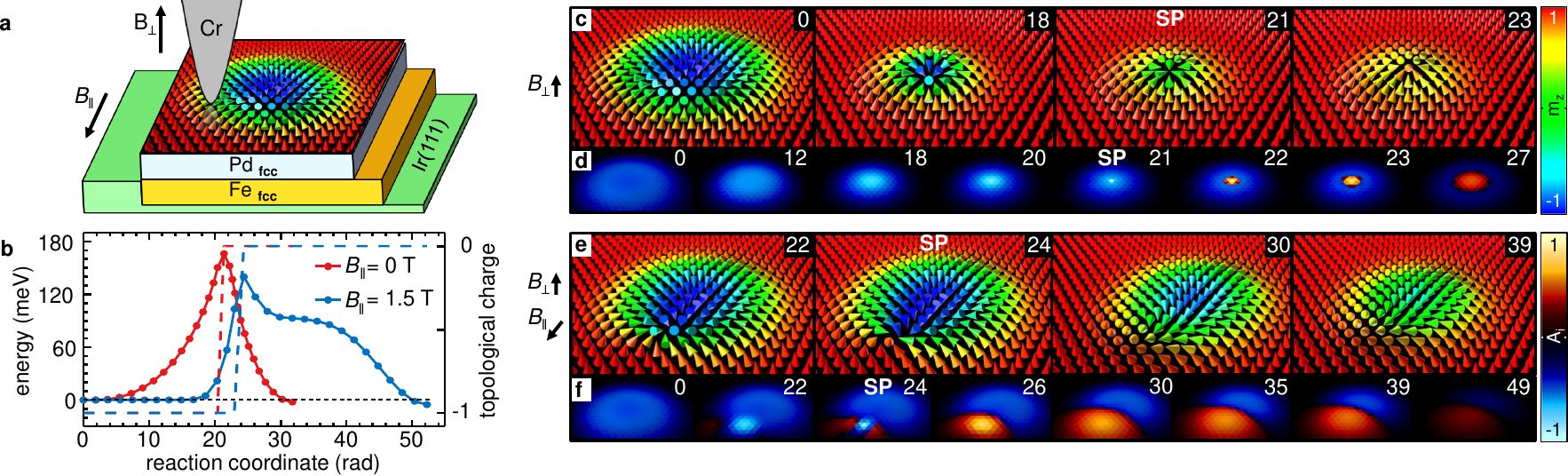}
\caption{\label{fig1}
\textbf{Radial symmetric and chimera collapse mechanism.}
\textbf{a}, Sketch of the experiment with indicated applied $B$ fields. \textbf{b}, %
Energy relative to the initial state (dots) and total topological charge $Q_{\rm topo}=\sum_i A_i$ 
(dashed lines, Methods) for images along the reaction coordinate of the radial symmetric collapse (red) at $B_\parallel=0$\,T and of the chimera type collapse (blue) at $B_\parallel=1.5$\,T deduced by GNEB, $B_\perp=3.2$\,T. \textbf{c} Selected calculated images of magnetization vectors along the GNEB minimum energy path shown in (b) for the radial symmetric skyrmion collapse. Cones mark the magnetization directions ${\bm m}$ for each atom. The color indicates the $z$ component of the magnetization $m_z$. Inset number indicates the reaction coordinate along the energy path as marked in (b). \textbf{d}, Maps of topological charge per triangle $A_i$ 
(Methods) %
along the same GNEB minimum energy path. %
\textbf{e, f}, Same as (c),(d) for the chimera collapse. More details on both mechanisms are shown in supplementary Fig. S15-S16 and movies 1-2.}
\end{figure*}
Here, we show experimentally that both, the radial symmetric and the chimera-type collapse indeed exist. %
We map the current induced skyrmion collapse rate of the model system Pd/Fe/Ir(111) with sub-nm resolution by SP-STM \cite{Bode2003}. %
We show that the collapse probability per injected electron exhibits a symmetric map with respect to the simultaneously imaged skyrmion center for zero in-plane field and becomes strongly asymmetric with applied in-plane fields. %
This is straightforwardly explained by the energy distributions required for the transition states of the radial symmetric and chimera type collapse, as calculated via an atomistic spin model parametrized by DFT results and using the geodesic nudged elastic band (GNEB) approach \cite{Bessarab2015} and transition state theory (TST) \cite{Bessarab2012}. %
We experimentally tune the collapse mechanism and rate by both, in plane and out-of-plane magnetic fields (Fig.~\ref{fig1}a), changing the rate by up to four orders of magnitude. %
We also establish that both collapse mechanisms are mostly induced by the local energy injection of a single hot electron. 
The fact that in-plane magnetic fields can strongly tune the skyrmion collapse might be exploited to improve skyrmion stability as well as for controlled skyrmion annihilation. %

\section*{Two skyrmion collapse modes}
Figure\,\ref{fig1}a sketches the experimental STM setup
to map magnetic skyrmions in a biatomic Pd/Fe bilayer, fcc stacked on the Ir(111) surface.
Depending on the applied in-plane, $B_\parallel$, and out-of-plane, $B_\perp$, magnetic fields, the skyrmion 
collapses via the radial symmetric or the chimera-type mechanism (Fig.\,\ref{fig1}b).
Spin configurations along the reaction coordinate
of the usual radial symmetric collapse, obtained via the GNEB method, are shown in Fig.\,\ref{fig1}c 
(more details: Supplementary movies).
The spin structure exhibits the well-known shrinkage of the area with spins that are not oriented along 
$B_\perp$ \cite{Bessarab2015, Rohart2016}. At the saddle point (SP), i.e.~the image along the reaction 
coordinate defining the energy barrier (Fig.\,\ref{fig1}b),
three central spins point towards each other representing the topology flip. Indeed, maps of the topological charge density
$\rho_{\rm{topo}}=\frac{1}{4\pi}{\bm m}\cdot (\partial_x {\bm m} \times  \partial_y {\bm m} )$ (${\bm m}={\bm M}/M$: 
normalized magnetization), as realized on a discrete lattice via the topological charge per triangle $A_i$ (Methods), exhibit an initial shrinkage towards this point (Fig.\,\ref{fig1}d).
Afterwards, the central part of $\rho_{\rm topo}$ reverses sign accompanying the sign change of the $z$ component of 
${\bm m}$, $m_z$, before annihilating with its surrounding. 
The energy  along the reaction coordinate 
exhibits a  maximum close to the topology flip (Fig.\,\ref{fig1}b).

The chimera process is displayed in Fig.\,\ref{fig1}e.
In contrast to the radial symmetric collapse, the spin structure at the SP has barely shrunk.
Instead, the spins opposing $B_\parallel$ have rotated in out-of-plane direction with the spins in the inner and the outer skyrmion area pointing oppositely. 
This gives rise to a Bloch-like point at the left foreground such that the canting between neighboring spins is strongly anisotropic. 
Subsequently, some spins of this area flip their in-plane direction and, thus, annihilate the Bloch-like point. This introduces an area of opposing $\rho_{\rm{topo}}$ that exactly cancels with the remaining topological charge of the rest of the spin structure. 
Afterwards, the spin structure does not require further flips while continuously rotating until being aligned completely with $B_\perp$.

\begin{figure}
\includegraphics[width=88mm]{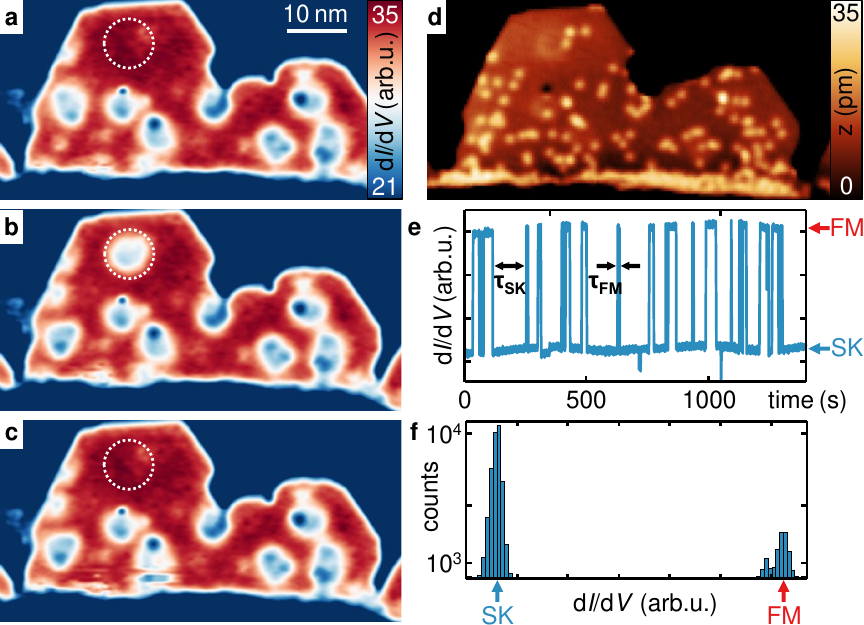}
\caption{\label{fig2}
\textbf{Current induced skyrmion switching (Pd/Fe/Ir(111)).}
\textbf{a-c}, Subsequently recorded $dI/dV$ maps of the same area  featuring the creation ((a) $\rightarrow$ (b)) and annihilation ((b) $\rightarrow$ (c)) of a single skyrmion, $I =  300$\,pA, $V=610$\,mV. 
\textbf{d}, Topographic image recorded simultaneously with (b). 
\textbf{e},  Time trace of $dI/dV$ signal recorded in the center of the dotted circle of (a)-(c), $I = 10$\,nA, $V=610$\,mV. Two individual life times of skyrmion ($\tau_{\rm SK}$) and ferromagnetic state ($\tau_{\rm FM}$) are marked. 
\textbf{f}, Histogram of $dI/dV$ values from the time trace of $\Delta t =1430$\,s, shown in (e) and exhibiting two distinct peaks that represent the skyrmion (SK) and the ferromagnetic (FM) state.
\textbf{a-f} $B_\parallel=0$\,T, $B_\perp=1.5$\,T, $T=6$\,K.}
\end{figure}

The barrier height of the two collapse mechanisms is similar at the chosen magnetic fields, while the shape of the minimum energy path is different (Fig.\,\ref{fig1}b). 
The shrinking of the skyrmion during the radial symmetric collapse continuously costs energy, while the chimera mode exhibits a lateral movement of the skyrmion center during the first part of the reaction path that barely requires energy. 
It is followed by a sharp energy maximum around the SP and a subsequent plateau marking the continuous spin rotation of the vanishing spin texture (more details: Supplementary section S8).
While the out-of-plane spin flip at the SP of the radial collapse is assisted by $B_{\perp}$, the in-plane spin flip at the SP of the chimera mode is favored by $B_{\parallel}$ and, thus, a transition between the two mechanisms is expected by tuning the magnetic fields. 

\begin{figure*}[ht]
\includegraphics[width=176mm]{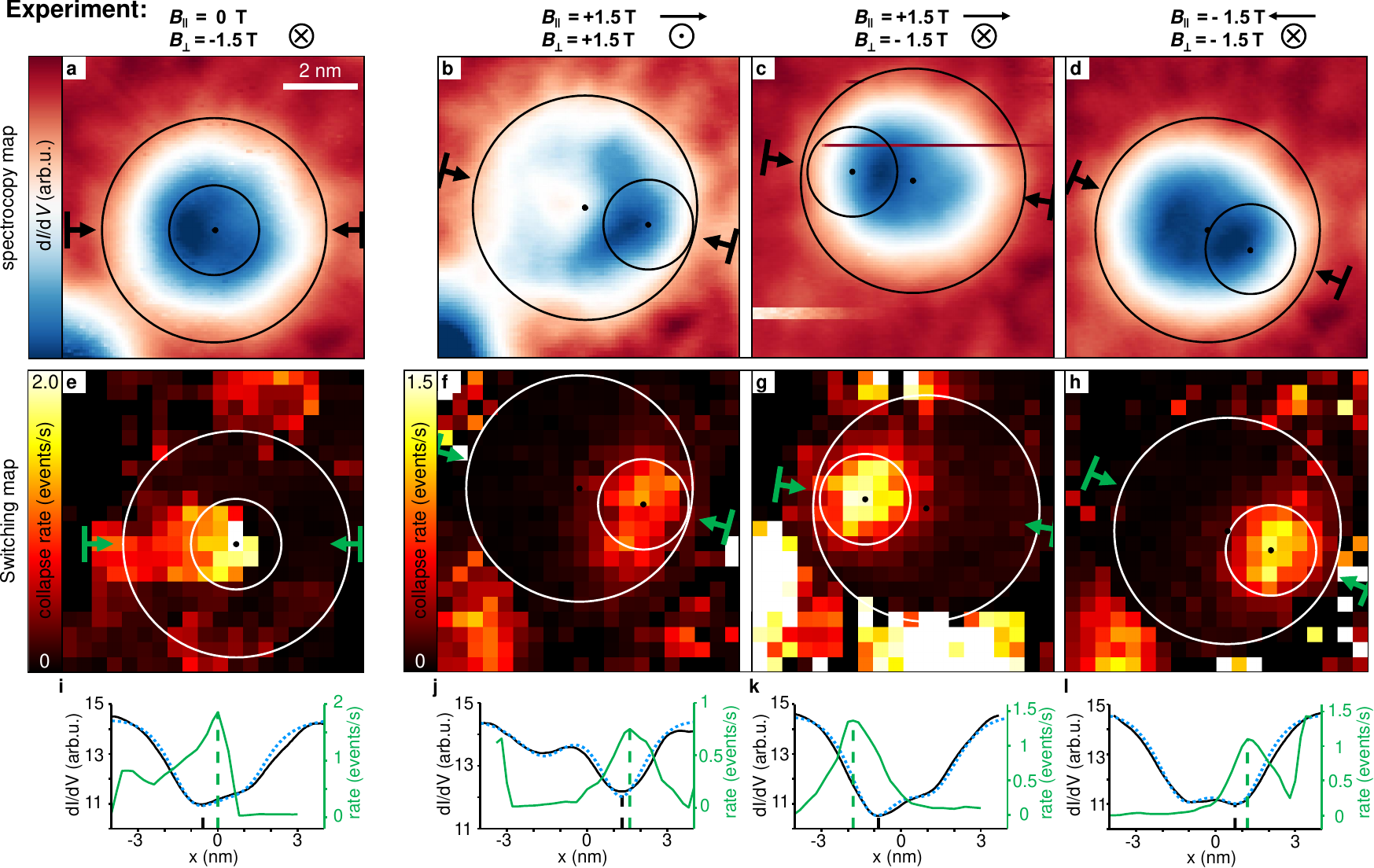}
\caption{\label{fig3}
\textbf{Maps of skyrmion collapse rate in Pd/Fe/Ir(111).}
\textbf{a-d}, Experimental $dI/dV$ maps of the same skyrmion at different magnetic fields ${\bm B}$ as indicated above the images, $I=1$\,nA,  $V=610$\,mV.
The larger circle encloses the skyrmion area with radius $r=3$\,nm as deduced by adapting the $dI/dV$ contrast to the spin canting such that the spins on the circle are canted by $\sim 10^\circ$ with respect to the out-of-plane direction (Methods, Supplementary section S3). The smaller circle with radius $r=1.2$\,nm encloses the area of increased collapse rate as determined in (e)$-$(h). Black dots mark the center of the circles.  Arrows indicate the orientation of profile lines displayed in (i)-(l). 
\textbf{e-h} Experimental maps of the skyrmion collapse rate corresponding to the dI/dV maps above. Each pixel ($20\times 20$ per map) results from a time trace of  $400$\,s at $I= 600$\,nA in (e) and $150$\,s at $I= 60$\,nA (f), $I= 50$\,nA (g) and $I= 55$\,nA in (h). Circles have identical radius and position as in (a)-(d) with the smaller circle roughly marking the FWHM of the hot spot. Arrows are at identical positions as in (a)-(d), too, and mark the directions of profile lines in (i)-(l). \textbf{i-l}, Profile lines of collapse rate (green) and $dI/dV$ contrast (black) along the directions marked by arrows in the corresponding images. The blue dotted lines are $dI/dV$ profiles of skyrmions as calculated from the spin textures of the DFT based atomistic spin simulations (Fig.~\ref{fig4}, Supplementary section S3). All profiles are averaged across 1 nm (widths marked by the thicker bars perpendicular to the arrows in the corresponding maps). Dashed vertical lines mark maximum (minimum) of the green (black) curve highlighting a small, but systematic mutual offset.}
\end{figure*}

\section*{Mapping skyrmion collapse rates}
Figure\,\ref{fig2} introduces the experimental access to the collapse mechanisms via mapping the skyrmion flip rate. 
Figure~\ref{fig2}a-c  show the same area several times as recorded by SP-STM at $B_\perp =1.5$\,T. A number of skyrmions is visible as roughly circular contrasts with rich internal structure. 
The contrast is caused by 
a spin-polarized contribution to the tunnel
current due to the magnetic tip \cite{Bode2003,Romming2015} 
and an electronic contribution due to the noncollinear arrangement of 
spins coined the non-collinear magnetoresistance (NCMR) \cite{Hanneken2015,Kubetzka2017}. The sequence of three images 
(Fig.~\ref{fig2}a-c) showcases the presence and absence of one skyrmion as marked by a white circle. This skyrmion covers an area of reduced defect density (Fig.\,\ref{fig2}d), while all other, more stable skyrmions are accompanied by several defects implying that defects enhance the skyrmion stability. Importantly, the switching rate of the largely free skyrmion increases with increasing tunnel current $I$
as discussed further below. This leads to the telegraph noise recorded at the white circle at larger $I$ (Fig.\,\ref{fig2}e), where the $dI/dV$ signal flips between two values (Fig.~\ref{fig2}f) representing the creation and annihilation of the skyrmion.  By  reducing $I$ at the lower $dI/dV$ value, the 
created skyrmion can be mapped as in Fig.~\ref{fig2}b,  while reducing $I$ at the upper $dI/dV$ state enables mapping of the ferromagnetic state without skyrmion as in Fig.~\ref{fig2}c.

We use this telegraph noise to determine the flip rate of the skyrmion as function of current injection point. This leads to flip rate maps for creation and collapse of the skyrmion separately. Figure \ref{fig3} e-h show the collapse rate maps at different ${\bm B}$ in comparison with the $dI/dV$ maps of the same skyrmion probed at low $I$ (Fig.\,\ref{fig3}a-d). Maps of the corresponding creation rates are largely homogeneous within the skyrmion area (Supplementary section S1). Most intriguingly, the maps of the collapse rate exhibit a hot spot within the area of the skyrmion, i.e., a small area with markedly increased rate (profile lines in Fig.\,\ref{fig3}i-l). This hot spot is located in the center of the skyrmion at $B_\parallel = 0$\,T (Fig.\,\ref{fig3}e) and displaced sidewards close to the rim of the skyrmion at finite $B_\parallel$ (Fig.\,\ref{fig3}b-d). The displacement direction flips sign by changing either the sign of $B_\perp$ (Fig.\,\ref{fig3}f $\rightarrow$ g) or the sign of $B_\parallel$ (Fig.\,\ref{fig3}g $\rightarrow$ h). Comparison with the $dI/dV$ maps (Fig.~\ref{fig3}a$-$d) reveals that the hot spot is always located close to the area of smallest $dI/dV$, marking the area of strongest spin canting between neighboring spins via the dominating NCMR contrast (Supplementary section S3). 
The collapse is, hence, strongly favored by injecting electrons at the area of strong spin canting, where spin flips are easier due to the initially strong exchange energy density. This area of strong spin canting is markedly displaced from the skyrmion center at finite $B_\parallel$, while the out-of-plane spins opposing $B_\perp$ are still located close to the center (Supplementary Fig. S13).
Inverting the direction of either $B_\perp$ or $B_\parallel$ switches the position of strongest relative spin canting with respect to the skyrmion center and, consistently, we observe a position switch of the hot spot in the skyrmion collapse rate.
Interestingly, the central hot spot at $B_\parallel = 0$\,T (Fig.\,\ref{fig3}e, i)  is accompanied by a small side arm to the left that points towards the rim of the island (skyrmion A in Fig.~\ref{fig6}a) as discussed in supplementary section S5.

\section*{Calculating skyrmion collapse paths}
In order to explain the experimental observations, we perform 
minimum energy path simulations 
based on an atomistic spin model 
parametrized from DFT calculations for Pd/Fe/Ir(111) (Methods). The minimum energy paths reveal which spins need to be activated so as to bring the skyrmion over the energy barrier and to trigger the collapse. The positions of those active spins can be compared with the location of the measured hot spots of the skyrmion collapse rates, thereby providing the interpretation of the experimental data.

\begin{figure}
\includegraphics[width=80mm]{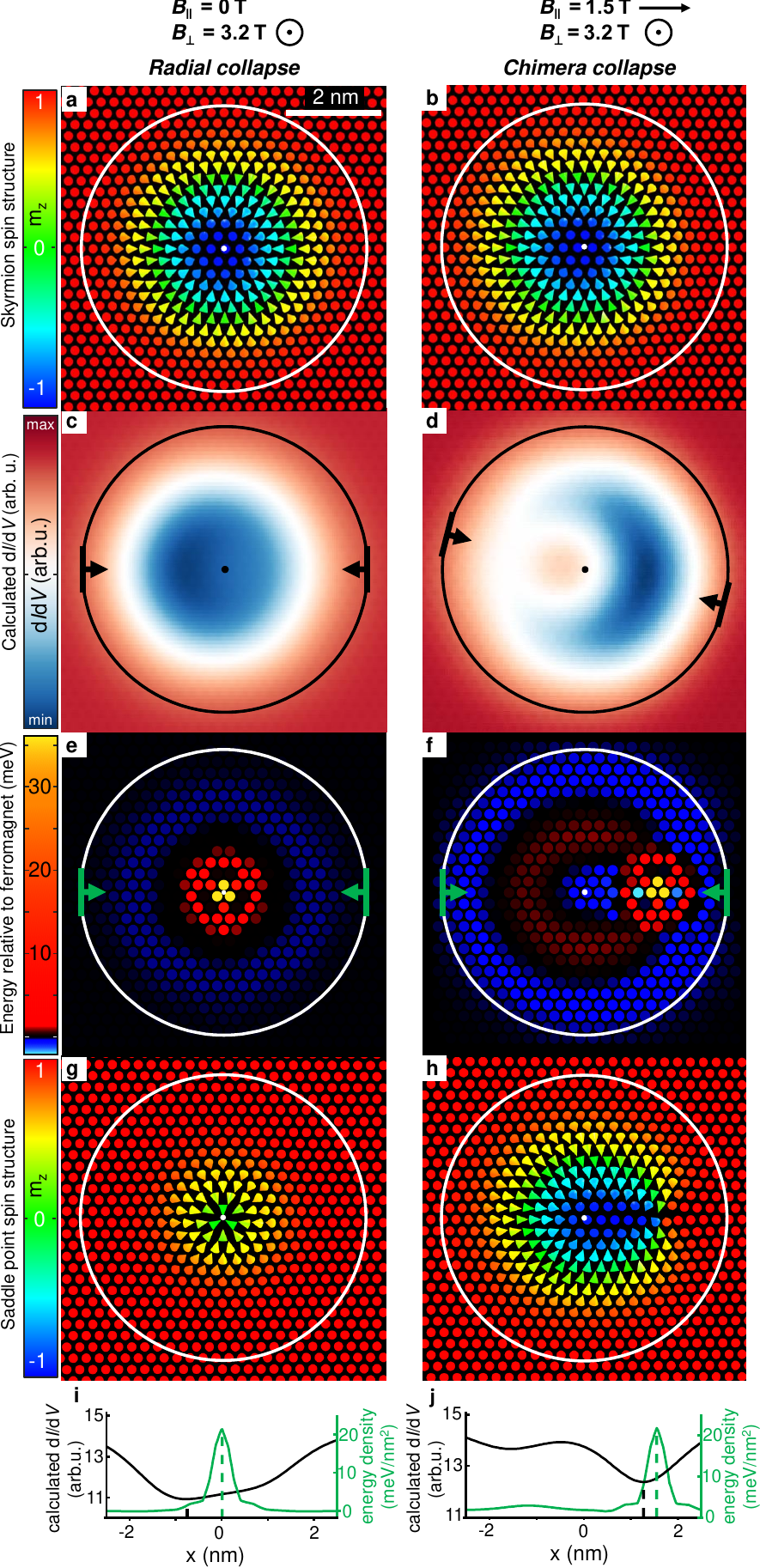}
\caption{\label{fig4}
\textbf{Simulated skyrmion collapses in Pd/Fe/Ir(111).} 
\textbf{a,b}, Spin structure of the relaxed skyrmion state at $B_\perp=3.2$\,T, (a) $B_\parallel = 0$\,T, (b) $B_\parallel = 1.5$\,T. 
\textbf{c,d}, Simulated $dI/dV$ maps of the spin structures shown in (a)$-$(b) assuming contributions of NCMR contrast and tunneling magnetoresistance via a spin-polarized STM tip (Supplementary section S3).
 \textbf{e,f}, Calculated energy density maps at the saddle point (SP) of the radial symmetric (e) and chimera (f) collapse. Arrows in (c)$-$(f) mark directions and interpolation widths of profile lines shown in (i)$-$(j).
\textbf{g,h}, Spin structures at SP for the radial (g) and  the chimera (h) collapse.
Circles in (a)$-$(h) have the same size ($r=3$\,nm) identical to the larger circles in Fig.~\ref{fig3}(a)$-$(h).
\textbf{i}, Profile lines along arrows in (c), (e), i.e., for the radial collapse.
\textbf{j}, same profile lines for (d), (f), i.e., for the chimera collapse. Black lines are identical to the blue dotted lines in Fig.~\ref{fig3}(i)$-$(j). 
Dashed lines mark maximum (minimum) of the green (black) curve for comparison with Fig.~\ref{fig3}(i)$-$(j).}
\end{figure}

Figure\,\ref{fig4} illustrates the skyrmion 
collapse  
without and with in-plane field $B_\parallel$.  
We use the skyrmion radius as benchmark for the comparison with the experiment to adapt $B_\perp$ \cite{vonMalottki2017}. It is deduced by matching measured and simulated $dI/dV$ profile lines (Fig.~\ref{fig3}(i)$-$(l)), eventually implying $B_\perp=3.2$~T in the simulations (Methods, Supplementary section S3). 
The spin structure of the skyrmion state is only little affected by $B_\parallel$ (Fig.\,\ref{fig4}a,b).
However, in agreement with experiment (Fig.~\ref{fig3}a,b), the simulated SP-STM images (Fig.\,\ref{fig4}c,d) exhibit a strong asymmetry that is traced back to an anisotropic canting angle of neighboring spins made visible via the dominating NCMR (Supplementary section S3). 

Without $B_\parallel$, the calculated skyrmion collapse is radially symmetric via a shrunk spin texture at the SP (Fig.~\ref{fig4}g). The energy density map at the SP, displaying the total energy per atom with respect to the ferromagnetic (FM) state, 
is radially symmetric with maximum at the center (Fig.~\ref{fig4}e). This illustrates the increased exchange energy due to the strong relative canting of the central spins at the SP. 
An energy gain on a ring around the center is also visible due to the stronger rotation of neighboring spins in this area leading to a gain in Dzyaloshinskii-Moriya interaction (DMI) relative to the FM state (Supplementary Fig.~S10).
At finite $B_\parallel$, the simulations reveal a 
transition to the chimera collapse showcasing an asymmetric spin configuration at the SP  with a Bloch-like
point at the right circumference area (Fig.~\ref{fig4}h). The size of the skyrmion barely changes up to the SP. The energy density map at the SP (Fig.~\ref{fig4}f) is asymmetric with a maximum at the Bloch-like point due to the increased exchange energy there.

Intriguingly, the hot spots of the calculated energy density maps at the SP (Fig.~\ref{fig4}e,f) match the measured hot spots of the skyrmion collapse rates (Fig.~\ref{fig3}e,f). This is our central result that experimentally  evidences a transition from the radial symmetric to the chimera-type collapse with increasing $B_\parallel$. In Supplementary section S9, we demonstrate that the radial collapse at finite $B_\parallel$ still has the hot spot of SP energy density close to the center of the skyrmion and not at the rim. 

\section*{Comparing collapse rates}
In the following, we will substantiate this central result by comparing the collapse rates between experiment and theory quantitatively and by providing a rough model of the skyrmion collapse via a single hot electron process. 

\begin{figure}[ht]
\includegraphics[width=88mm]{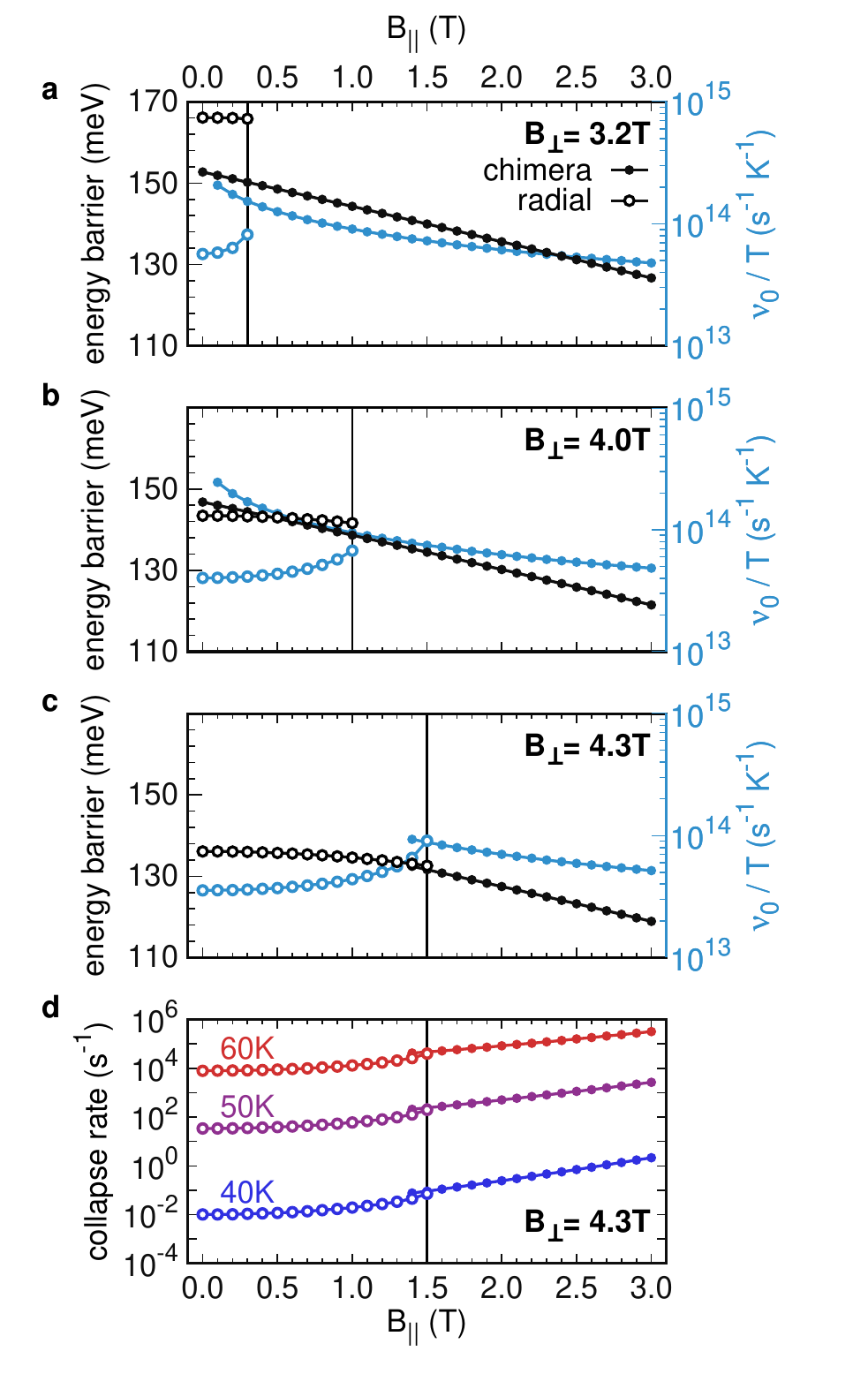}
\caption{\label{fig5}
\textbf{Calculated energy barriers and prefactors for skyrmion collapse (Pd/Fe/Ir(111)).}
\textbf{a-c}, Energy barriers (black) and prefactors divided by temperature, $\nu_0/T$ (blue), for the radial symmetric (open symbols) and chimera collapse (full symbols) of an isolated skyrmion at varying $B_\parallel$ according to atomistic spin simulations via GNEB and TST based on parameters from DFT (Methods). (a) $B_{\perp}=3.2$\,T, (b) $B_{\perp}=4.0$\,T, (c) $B_{\perp}=4.3$\,T.
\textbf{d}, Thermal collapse rates resulting from (c) using eq.~(\ref{eq:Arrhenius}) for three temperatures with filled (open) circles employing the chimera (radial symmetric) collapse. Vertical lines in (a)$-$(d) mark the maximum $B_\parallel$, at which the radial symmetric collapse could be stabilized.}
\end{figure}

Theoretically, we obtain the collapse rates $\nu$ within quasi-equilibrium implying an Arrhenius law at fixed temperature $T$:
\begin{equation}
\nu = \nu_0 \exp{\left(-\frac{\Delta E}{k_{\rm B}T}\right)}.
\label{eq:Arrhenius}
\end{equation}
Here, $\nu_0$ is the attempt frequency, also called prefactor, and $\Delta E$ is the energy barrier. Both quantities depend on the mechanism of skyrmion collapse and are calculated  within harmonic approximation to the rate theory as a function of $B_\parallel$ for various $B_\perp$ (Fig.~\ref{fig5}) (Methods).
At $B_\perp = 3.2$\,T (Fig.~\ref{fig5}a), the radial collapse mechanism occurs only at very small $B_\parallel<0.3$~T and
exhibits a higher energy barrier (black circles) than the chimera collapse (black dots). The
chimera collapse occurs at all $B_\parallel$ with energy barrier decreasing 
with increasing $B_\parallel$ due to the increased spin canting in areas where magnetization opposes $B_\parallel$. Hence, the spin flips get easier in these areas. The prefactor of the chimera collapse (blue dots, Fig.\,\ref{fig5}a) decreases with $B_\parallel$, too, and is similar in magnitude to the prefactor of the radial collapse (blue circles) at low $B_\parallel$.
At $B_\perp=4.0$\,T (Fig.~\ref{fig5}b), the radial collapse is stable up to  $B_\parallel=1$\,T with energy barrier 
partially lower than for the chimera mode up to a crossing point at $B_\parallel \simeq 0.5$\,T. The prefactor of both mechanisms increases in direction of the energetic crossover due to mode softening that increases the entropy of the transition states  \cite{Bessarab2013}.
Far from the crossover, the prefactors of both mechanisms approach
a similar value. 
At even higher $B_\perp=4.3$\,T (Fig.~\ref{fig5}c), the radial symmetric collapse is favorable up to      
$B_\parallel=1.5$\,T, where the chimera collapse is unstable, while the opposite is true at larger $B_\parallel$. 

These results at various $\bm B$ demonstrate consistently that the 
symmetric collapse is favored by $B_\perp$, while $B_\parallel$ supports the chimera-type mechanism, basically by mutually tuning the positions of strongest relative spin canting between center and rim.
Based on the calculated prefactors and energy barriers, eq.(\ref{eq:Arrhenius}) provides collapse rates $\nu(B_\parallel)$ at different $T$ as shown for $B_\perp=4.3$\,T (Fig.~\ref{fig5}d). 
As expected, the rate increases drastically with $T$ and rises with $B_\parallel$ exhibiting a transition from preferential radial collapse to preferential chimera-type collapse for $B_\parallel \simeq 1.5$~T.

\begin{figure*}[ht]
\includegraphics[width=176mm]{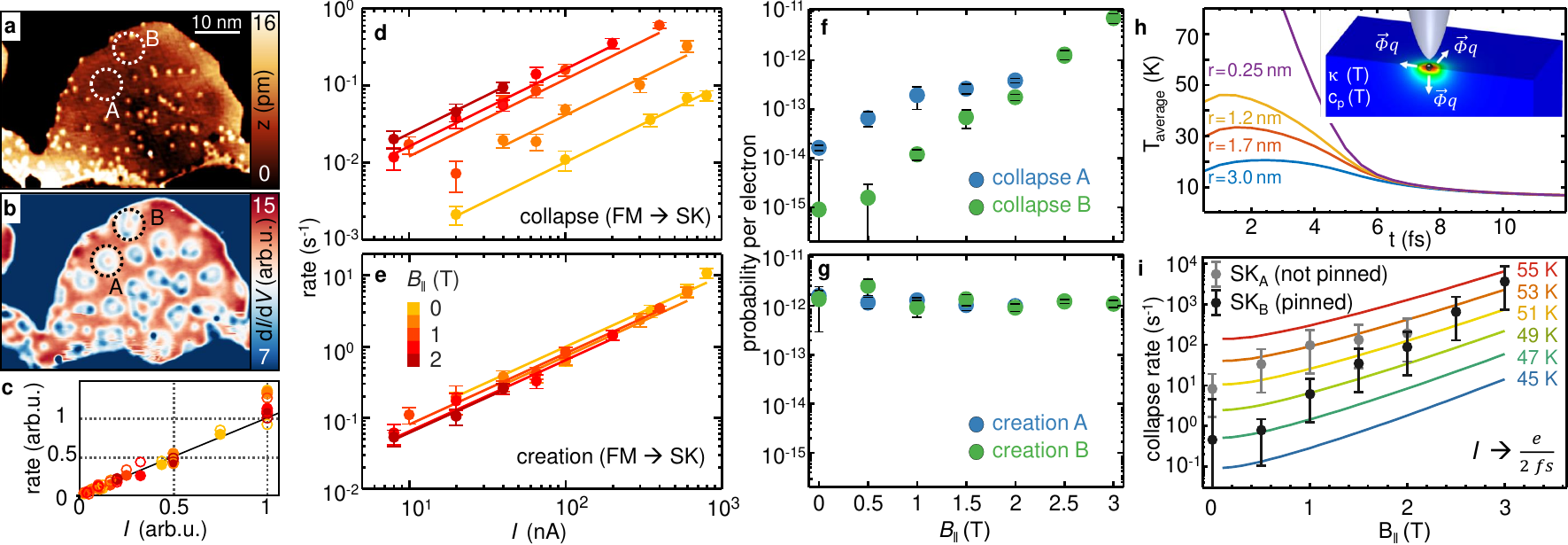}
\caption{\label{fig6}
\textbf{ Skyrmion collapse and creation rates at different $B_\parallel$ (Pd/Fe/Ir(111)).}
\textbf{a} STM topography of a Pd island on Fe/Ir(111) with the two investigated skyrmions encircled, $B_{\parallel}=0$\,T, $B_{\perp}=1$\,T, $V=610$\,mV, $I= 2$\,nA. \textbf{b} Simultaneously recorded $dI/dV$ map with the same skyrmions encircled and labeled as in (a). \textbf{c} Collapse rates (circles) and creation rates (dots) as function of tunnel current $I$ at different $B_\parallel$ (color code in (e)).
Both, $I$ and rate $\nu$ are scaled to exhibit the same slope $\Delta \nu/\Delta I$ of the linear fit curve to the data to display all data within a single graph. The linearity implies that the collapse is induced by independent single electron events. 
\textbf{d} Collapse rate of skyrmion A 
as function of $I$ in log-log scale at different $B_\parallel$ marked via symbol colors according to legend in (e), $V=610$\,mV, $B_{\perp}=1.5$\,T. Linear fit lines are added in identical color. \textbf{e}, Creation rate of skyrmion A  at the same parameters as in (d).
\textbf{f, g}, Skyrmion collapse (f) and creation (g) probability per electron as deduced from linear fits of (d), (e) for the two skyrmions marked in (a), (b), $B_{\perp}=1.5$\,T, $V=610$\,mV. Collapse and creation rates of skyrmion B are shown in Supplementary Fig.~S3. \textbf{h} Simulated temporal development of mean temperature within a surface area of radius $r$, after inducing the energy of a single electron ($V=610$ mV) homogeneously within $r=0.25$\,nm at $t=0$\,fs. Inset: Sketch of the finite element model with indicated heat flow $\vec{\Phi}_q$, heat  conductivity $\kappa(T)$, and heat capacitance $c_p(T)$.
\textbf{i} Calculated thermal collapse rates at different $T$ using the theoretical parameters from Fig.\,\ref{fig5}a (colored lines) and compared to the experimental single electron probabilities (f) that are extrapolated to collapse rates for an assumed  $I= e/2$\,fs~$=75{\rm \mu}$A (symbols) such that the area of $r=1.2$\,nm (hot spot in Fig.\,\ref{fig3}) would be continuously heated to  $\sim 45$\,K according to the calculation of (h).}
\end{figure*}

Experimentally, we obtain the trend of skyrmion collapse rates with $B_\parallel$ by measuring the random telegraph noise  within the central skyrmion area. Again, we select a relatively defect-free skyrmion (skyrmion A in Fig.~\ref{fig6}a,b) and compare it to another skyrmion B that exhibits several 
defects in its area, known to pin the skyrmion \cite{Hanneken2016}.
Firstly, we show that the skyrmion collapse and creation are single electron processes, i.e., the corresponding switch rates scale linearly with $I$  (Fig.\,\ref{fig6}c). This applies for, both, skyrmion collapse and creation at all studied magnetic fields 
(Fig.~\ref{fig6}d$-$e), hence, enabling to deduce a probability for collapse and creation per injected electron (Fig.\,\ref{fig6}f-g). The resulting collapse probability $P_{\rm collapse}$ increases by two orders of magnitude with $B_\parallel$ for the defect-free skyrmion A (Fig.~\ref{fig6}f), prior to its disappearance at larger $B_\parallel$. This fits nicely with the change in collapse rate by about two orders of magnitude found theoretically
 at constant $T$ (Fig.~\ref{fig5}d). The pinned skyrmion B is stable up to larger $B_\parallel$ and its $P_{\rm collapse}(B_\parallel)$ covers even four orders of magnitude. Moreover, $P_{\rm collapse} (B_\parallel)$ is significantly steeper for skyrmion B, highlighting the importance of defects that tend to stabilize skyrmions in Pd/Fe/Ir(111) as corroborated by probing multiple skyrmions on that surface. In contrast, the creation probability (Fig.~\ref{fig6}g) barely depends on $B_\parallel$ as also found theoretically via the atomistic spin simulations (Supplementary Fig.~S14 and S15). It also barely depends on the defect configuration.
 
Eventually, we compare the measured $P_{\rm collapse} (B_\parallel)$ (Fig.~\ref{fig6}f) with the calculated collapse rates at constant $T$. For this purpose, we use a strongly simplified model assuming that each electron heats a certain area of the skyrmion for a short time such that the collapse can proceed via eq.~\ref{eq:Arrhenius}. We employ a finite element calculation for a  semi-infinite crystal and initially deposit the energy of a single hot electron in a small circle (radius: 0.25\,nm) at the surface.  Subsequently, we calculate the 3D propagation of temperature $T$ as function of time $t$ using heat capacitance $c_p(T)$ and heat conductivity $\kappa(T)$ of Pd (Methods, inset of Fig.~\ref{fig6}h).      
Figure\,\ref{fig6}h displays the resulting spatially averaged $T(t)$ for different areas, namely the area of energy deposition ($r=0.25$\,nm), the area of the skyrmion ($r=3.0$\,nm), the area of the hot spot for collapse ($r=1.2$\,nm, Fig. \ref{fig3}) and a radius in between ($r=1.7$\,nm). 

The whole skyrmion (hot spot) is heated by each electron up to $\sim 20$\,K ($\sim 45$\,K) for a few fs.
This time is much shorter than the average time $\tau$ between sequential electrons of the tunnel current $\tau = e/I \ge 150$\,fs ($e$: electron charge). Such time scale mismatch straightforwardly explains the deduced single electron processes, i.e., the skyrmion is always cooled down prior to the next electron injection. 
A rough estimate of the heat induced dynamics is given by considering the hot spot area to be at $T \simeq 45$\,K for $\Delta t\simeq 2$\,fs after injecting the electron, respectively, the whole skyrmion area at $T \simeq 20$\,K for $\Delta t\simeq 4$\,fs (details of the estimate: Supplementary section S7). 

In turn, constantly increased $T$ within these areas implies a current $I=e/\Delta t$ carried by independent electrons. Accordingly, we scaled the experimental $P_{\rm collapse} (B_\parallel)$ to $I=e/\Delta t$ obtaining a collapse rate at quasi-constant $T$. These quasi-experimental rates can be straightforwardly compared to the theoretically deduced rates from Fig.~\ref{fig5}a. Figure~\ref{fig6}i provides such comparison for $\Delta t = 2$\,fs. Intriguingly, the experimental data of the defect-free skyrmion A (grey symbols) match the theoretical data for an average temperature of $T\simeq 50$\,K, very close to the estimated $T\simeq 45$\,K of the hot spot. 

Thus, shortly heating the hot spot via a single hot electron to $\sim 50$\,K appears to be the central ingredient to the skyrmion collapse. The additional spin torque of each hot electron sincerely contributes, but this influence is less crucial, i.e., mostly part of the Arrhenius type statistics. Doubling $\Delta t$, i.e., adapting the scaled $I$ in Fig.~\ref{fig6}i towards heating the whole skyrmion to a continuous temperature, does not significantly change the required $T \sim 50$\,K for matching theory. Consequently, heating the hot spot is more crucial than heating the whole skyrmion that only achieves $T\simeq 20$\,K. 

This straightforwardly implies that the rate must significantly increase, if the hot electron hits the hot spot of energy density at the SP directly, hence, explaining the observed hot spots in collapse rates of Fig.~\ref{fig3}. The  very good agreement of temperatures between theory and experiment strongly corroborates our interpretation that the measured off-center hot spot in the collapse rate is linked to the calculated hot spot in SP energy density of the chimera collapse. Moreover, the single hot electron mechanism explains why previous Monte-Carlo simulations had to assume $T=80$\,K to fit the experimental collapse data observed at $T=4$\,K \cite{Hagemeister2015}.

\section*{Conclusions and Outlook}
Mapping transition rates that are initiated by local energy deposition on the sub-nm scale enabled us to determine the typically elusive transient state. This gives direct access to the transition mechanism that, so far, is mostly only inferred from parameter-based simulations. In the particular case, the unwinding of the spin texture of a skyrmion, largely protected by topology, turns out to exhibit two competing mechanisms that are quite distinct regarding its transient state, but nevertheless, can be tuned towards dominating by subtle parameter changes such as by an in-plane magnetic field. The resulting detailed understanding of the transient state, that limits the stability, is crucial to prohibit annihilation of information carriers such as skyrmions \cite{Fert2017,Jiang2015,Legrand2017}. Moreover, the knowledge can be employed to design deleting of information by employing the strong dependence of the collapse rate 
on $B_\parallel$. Hence, our experimental access  to transient states provides more reliable design criteria for  exploitation of the prospective, rather stable magnetic skyrmions. 

\begin{acknowledgments}
We gratefully acknowledge helpful discussions with S. Lounis, S. Bl\"ugel, A. Schlenhoff, M. Liebmann, M. A. Goerzen, T. Sigurj\'onsd\'ottir and financial support of the German Science Foundation (DFG)  via PR 1098/1-1, the Russian Science Foundation (Grant No. 19-72-10138), the Icelandic Research Fund (Grant No. 184949-052), and the Alexander von Humboldt Foundation.
\end{acknowledgments}


\section{Methods}
\textbf{Preparation of Pd/Fe bilayer.} The Ir(111) crystal was cleaned in ultra high vacuum (UHV) at base pressure $10^{-10}$\,mbar by repeated cycles of annealing up to $1200$\,K in oxygen at decreasing partial pressure from $10^{-6}$\,mbar to $10^{-8}$ mbar.  Additionally, cycles of argon ion bombardment at room temperature followed by flash annealing to $1700$\,K were performed. Subsequently, a monolayer (ML) of Fe was deposited by electron beam evaporation at sample temperature $470$\,K implying step-flow growth. Finally, 0.5 ML Pd were deposited by electron beam evaporation at substrate temperature of $400$\,K. For all measurements in this manuscript, we selected islands with fcc stacking via their characteristic dI/dV spectra \cite{Kubetzka2017} (Supplementary section S6). 

\textbf{Spin polarized STM.}
The tunneling tip is fabricated from a $0.5\times0.5$\,mm$^2$ beam of polycrystalline, antiferromagnetic Cr (purity 99.99+\%). Tip sharpening employs electrochemical etching by a suspended film of 2.5\,M NaOH solution within a PtIr loop held at 5.5\,V with respect to the tip. Etching is stopped at drop off of the lower beam part via differential current detection. The upper part of the beam is immediately rinsed with deionized water and glued onto a custom-made tip holder. The tip is then loaded into the UHV system and, subsequently, into the STM scan head at 6\,K \cite{mashoff2009}.
The atomic structure of the tip is optimized during tunneling by  voltage pulses (10\,V/30\,ms) between tip and sample until spin contrast is achieved. Voltage $V$ is applied to the sample. The differential conductance d$I$/d$V$ is measured by adding a 50\,mV RMS sinusoidal voltage (1384\,Hz) to the applied DC voltage $V$ and recording the resulting oscillation amplitude of the tunnel current $I$ using a lock-in amplifier. The system enables a 3D magnetic field ${\bm B}=(B_x,B_y,B_\perp)$ with out-of-plane component $B_\perp$ up to 7\,T and simultaneous in-plane part ${\bm B}_\parallel =(B_x,B_y)$ up to 1\,T in each in-plane direction and up to $3$\,T in a single direction  \cite{mashoff2009}.

\textbf{Determining switching rates.}
To determine the switching rates as displayed in Fig.~\ref{fig3}e$-$h and Fig.~\ref{fig6}c$-$e, we use time traces of $dI/dV$ values recorded in constant-current mode as exemplary shown in Fig.~\ref{fig2}e (more time traces: Supplementary Fig.~S5). For each time trace, we determine all $\tau_{\rm SK}$ and $\tau_{\rm FM}$ as dwell times of the corresponding $dI/dV$ level.
The collapse (creation) rate for a time trace is then calculated as the inverse of the average of the observed $\tau_{\rm SK}$ ($\tau_{\rm FM}$) dubbed $\overline{\tau}_{\rm SK}$ ($\overline{\tau}_{\rm FM}$). The skyrmion probability of the time trace reads:
\begin{equation}
  P_{\rm SK}=\frac{\overline{\tau}_{\rm SK}}{\overline{\tau}_{\rm SK}+\overline{\tau}_{\rm FM}}.
\end{equation}
Maps of switching rates (Fig.~\ref{fig3}e$-$h) are based on time traces between 150\,s and 400\,s at each position. 
Graphs of switching rates (Fig.~\ref{fig6}c$-$e) are based on time traces that last several 100\,s at each $I$ and $\bm B$.

\textbf{Atomistic spin model.}
The atomistic spin dynamics, GNEB, and TST calculations are performed in a simulation box of a hexagonal $70 \times 70$ 
atomic lattice with periodic boundary conditions. 
The applied extended Heisenberg Hamiltonian reads %
\begin{align}%
\mathcal{H}=&-\sum_{ij} J_{ij} \Big( \mathbf{m}_i \cdot \mathbf{m}_j \Big)  - \sum_{ij} \mathbf{D}_{ij} \Big( \mathbf{m}_i \times \mathbf{m}_j \Big) \notag\\ 
&+ K \sum_{i} \Big( \mathbf{m}_i^z  \Big) ^2 - M \sum_{i} \Big( \mathbf{m}_{i} \cdot \mathbf{B}_{\mathrm{ext}} \Big) %
\label{eq:spin_model}
\end{align}%
Here, $\mathbf{m}_i=\frac{\mathbf{M}_i}{M}$ is the normalized vector of the magnetic moment of the atom at the lattice site $i$. %
The parameters $J_{ij}$ and the vectors $\mathbf{D}_{ij}$ denote the strength of the exchange interaction and the strength and rotational sense of the DMI between the spins at $i$ and $j$, respectively. 
The strength of the uniaxial magnetocrystalline anisotropy in out-of-plane direction is given by the parameter $K$, while the external magnetic field $\mathbf{B}_{\mathrm{ext}}$ enters the Zeeman interaction. 
All parameters of eq.~(\ref{eq:spin_model}) 
including the magnetic moment $M$ are based on DFT calculations for fcc-Pd/fcc-Fe/Ir(111)
and taken from Ref.~\cite{vonMalottki2017}. %

\textbf{Selection of $B_\perp$ in the simulations and extraction of skyrmion radius.}
The DFT calculations provide a good quantitative description of the magnetic interactions for Pd/Fe/Ir(111) \cite{vonMalottki2017}.
In particular, the frustration of exchange interactions is captured which plays an important role for the 
occurrence of the chimera 
collapse \cite{Meyer2019,Heil2019}. However, there are deviations with respect to the
magnetic fields due to the relatively small Zeeman energy, i.e., the phase transitions between the spin spiral, the skyrmion lattice, and the field-polarized (ferromagnetic) 
state are shifted with respect to 
experiments~\cite{vonMalottki2017}. 
In order to select the adequate  $B_\perp$ fields for comparison with our experiments, we used the skyrmion size as benchmark.

 As shown in Fig.~\ref{fig3}(i)$-$(l), we could adapt the calculated $dI/dV$ profiles to the measured ones rather precisely using manual optimization (Supplementary Fig.~S4). It turned out that an increased $B_\perp$ in the atomistic spin simulations is sufficient to adapt the profiles, while $B_\parallel$ could be left as in the experiment. We used the resulting spin configurations of the calculations to determine the radius $r$ of the skyrmion defining the skyrmion circumference by the spins that are tilted by $\sim 10^\circ$ relative to $B_\perp$.
Application of $B_\parallel$ did not change $r$ via this definition, but shifted the out-of-plane spins within the circle by about one lattice site. Since the shape of the circumference additionally gets slightly elliptic with $B_\parallel$, we kept the radius and center position of the skyrmion from the determination at $B_\parallel=0$\,T, for the sake of simplicity. This simplification is irrelevant for our conclusions.    

\textbf{Calculation of minimum energy paths.}
The geodesic nudged elastic band method (GNEB) relaxes an initial path by following the energy gradient into a local minimum energy path (MEP) \cite{Bessarab2015}. %
A path consists of a discrete chain of states, so-called images, that interpolate between the isolated skyrmion, that is relaxed by spin dynamics simulations and the velocity projection optimization method \cite{Bessarab2015},
and the ferromagnetic state, chosen as initial and final state of the path, respectively. %
The images are connected by artificial spring forces so as to control the distribution along the path. %
The strength of the spring force for image $k$ caused by image $l$ depends on the geodesic distance in the configuration space $D_{\mathrm{geo}}$ and energy difference $D_{\mathrm{en}}$ between adjacent images via %
\begin{equation}
F_{\mathrm{spring}}^{k,l}=\kappa \sqrt{\left(D_{\mathrm{geo}}^{k,l}\right)^2+\kappa_{\mathrm{en}}\left(D_{\mathrm{en}}^{k,l}\right)^2}.
\end{equation}
Here, the spring constant $\kappa$ is chosen to be $0.1$ meV/rad$^2$. The measures of distance $D_{\mathrm{geo}}$ and $D_{\mathrm{en}}$ are in units of rad and meV and their relative weight is controlled by the parameter $\kappa_{\mathrm{en}}=0.01$ rad$^2$/meV$^2$. %
The extension of the method by taking $D_{\mathrm{en}}$ into account is necessary to obtain a sufficient number of images close to the SP of the chimera transition mechanism. This ensures a better resolution and convergence behavior of the method. %

In order to obtain metastable paths, first, an initial path is created by geodesic rotation from the initial to the final state that is additionally disturbed by small random fluctuations to avoid a path that is too symmetric to form an energy gradient. %
The GNEB relaxation of the geodesic path tends to converge into the radial symmetric annihilation mechanism for small $B_{\parallel}$ and large $B_{\perp}$, while it converges more often into the chimera type mechanism vice versa. %
Starting from these MEP configurations, $B_{\parallel}$ is successively decreased (increased) for the radial symmetric (chimera) mechanism, yielding relaxed metastable MEPs with varying $B_{\parallel}$ for both collapse mechanisms. %
Finally, a climbing image method is used to find the saddle point of the MEP. Different energetic contributions to the MEP and the SP are discussed in Supplementary section S8.%

\textbf{Calculation of skyrmion collapse rates.}
The rate of thermally activated transitions from the isolated skyrmion state to the ferromagnetic state is calculated in the framework of harmonic transition state theory, and therefore has the form of an Arrhenius law 
\cite{Bessarab2012,Bessarab2018}, i.e.~Eq.~(\ref{eq:Arrhenius}). 
The energy barrier $\Delta E$ of a MEP is provided by the energy difference between the saddle point (SP) and the initial state. 
By representing excitations of a state in harmonic approximation and in the eigenbasis of the corresponding Hessian, the prefactor $\nu_0$ reads: 
\begin{equation}
\nu_0 = \eta\frac{\lambda}{V}k_{\mathrm{B}}T\frac{\sideset{_{}^{}}{_{}^{}}\prod_i \sqrt{\epsilon_{\mathrm{SK},i}}}{\sideset{_{}^{}}{_{}^\prime}\prod_i \sqrt{\epsilon_{\mathrm{SP},i}}}. \label{eq:pref}
\end{equation}
Here, 
$\epsilon_i$ are the eigenvalues of the Hessian and $\lambda$ is the dynamical factor defining the system's velocity along the unstable mode.
The prime indicates that the negative eigenvalue corresponding to the unstable mode is excluded from the product. .
For the skyrmion state, two translational modes exhibit a sufficiently constant energy to be treated as Goldstone modes. They enter the prefactor via the volume $V$ per unit cell~\cite{Haldar2018}. In contrast, in-plane translations of magnetic structure at the SP corresponding to both the radial symmetric collapse and chimera mechanism are not energetically degenerate due to the presence of a defect whose size is comparable with the lattice constant. Thus, the corresponding modes are treated in harmonic approximation. For both collapse mechanisms, there are two equivalent SPs per unit cell as taken into account by a factor $\eta=2$ in Eq.~(\ref{eq:pref}). Unequal numbers of the Goldstone modes at the SP and at the skyrmion state lead to the linear temperature dependence of the prefactor. Additionally, motion of the Bloch point-like defect along the skyrmion's circumference in the chimera collapse shows a potential Goldstone behavior at zero $B_{\parallel}$. However, application of the in-plane magnetic field breaks this symmetry. As a consequence, the corresponding mode is also treated in harmonic approximation in our calculations. 

\textbf{Calculation of the topological charge}
We implemented the topological charge $Q_{\rm topo}$ on a discrete lattice \cite{muller2019} reading %
\begin{equation} %
Q_{\rm topo}= \sum_i A_i, %
\end{equation} %
where $A_i$ is the topological charge per triangle formed by three adjacent magnetic moments $\mathbf{m}_j,\mathbf{m}_k,\mathbf{m}_l$ with $i$ running over all triangles of the hexagonal lattice. 
The charge per triangle is calculated by \cite{muller2019} %
\begin{equation}%
\cos{\left(2 \pi A_i\right)}=\frac{1+\mathbf{m}_j \cdot \mathbf{m}_k + \mathbf{m}_j \cdot \mathbf{m}_l + \mathbf{m}_k \cdot \mathbf{m}_l}{\sqrt{2\left(1+\mathbf{m}_j \cdot \mathbf{m}_k \right)\left(1+\mathbf{m}_j \cdot \mathbf{m}_l \right)\left(1+\mathbf{m}_k \cdot \mathbf{m}_l \right)}}, %
\end{equation}%
with $\mathrm{sgn}\left(A_i\right)=\mathrm{sgn}\left[\mathbf{m}_j \cdot \left(\mathbf{m}_k \times \mathbf{m}_l\right)\right]$.%

\textbf{Calculation of temperature profiles after hot electron injection.}

We employed a finite element method as implemented in Solid works using the 
temperature dependent heat capacity $c_p(T)$ and heat conductivity $\kappa(T)$ of Pd within a virtually semi-infinite slab ($20\times 20 \times 15$ nm$^3$).
The starting configuration deposits the energy of a single hot electron (0.61\,eV) in a disk of radius 0.25\,nm and height of one atomic layer that is surrounded by a temperature of 6\,K. We use a grid size of half an atomic distance and time steps of 0.5\,fs for the calculations keeping all boundary cells except the ones at the surface at $T=6$\,K. For the sake of simplicity, we employ  $\kappa(T)$ and  $c_p(T)$ including the phonon contribution, albeit equilibration times between hot electrons and the phonon bath are typically longer than  a few fs \cite{ElsayedAli1987,Groeneveld1995}. This simplification keeps the parameters of the calculation as few as possible, in particular, concerning the complex simulation of heat conduction of hot electrons.  However, we keep in mind that the resulting temperature distribution remains a rough estimate, partly justified, since compared with theoretical calculations that assume an equilibrium temperature across the whole area of the calculation distinct from the regarded dynamics around a local hot spot only (more details: Supplementary section S7).   

\textbf{Data availability.}
The data that supports the plots within this paper and other findings of this study are available from the corresponding author upon request.

\bibliography{bibliography}
 
 \section{Author contributions}
 F. M. and B. P. conducted the experiments supervised by C. H., M. P. and M. M.. S. v. M. performed all atomistic spin simulations 
 supervised by P. F. B. and S. H.. F. M. and M. M. provided the original idea of the experiment with P. F. B.  and S. v. M. coming up with the idea of a more detailed comparison with DFT based calculations. In particular, S. v. M., P. F. B., and S. H. introduced the chimera mode as an explanation of the experimental data. F. M. with the help of C. H. evaluated all experimental data, performed comparisons to calculated data and conducted the calculations of heat distribution of a hot electron. M. M. wrote the first version of the manuscript with image supply by F. M. and S. v. M.. All authors contributed to multiple discussions and the final version of the manuscript.

\end{document}